\documentclass{vgtc}                          % final (conference style)
\ifpdf%                                % if we use pdflatex
  \pdfoutput=1\relax                   % create PDFs from pdfLaTeX
  \usepackage{graphicx}                % allow us to embed graphics files
  \DeclareGraphicsExtensions{.pdf,.png,.jpg,.jpeg} % for pdflatex we expect .pdf, .png, or .jpg files
\else%                                 % else we use pure latex
  \usepackage{graphicx}                % allow us to embed graphics files
  \DeclareGraphicsExtensions{.eps}     % for pure latex we expect eps files
\fi%

\graphicspath{{figures/}{pictures/}{images/}{./}} % where to search for the images

\usepackage{microtype}                 % use micro-typography (slightly more compact, better to read)
\PassOptionsToPackage{warn}{textcomp}  % to address font issues with \textrightarrow
\usepackage{textcomp}                  % use better special symbols
\usepackage{mathptmx}                  % use matching math font
\usepackage{times}                     % we use Times as the main font
\usepackage{cite}                      % needed to automatically sort the references
\usepackage{tabu}                      % only used for the table example
\usepackage{booktabs}                  % only used for the table example
\usepackage[dvipsnames]{xcolor}

\usepackage{color}
\definecolor{mscolor}{rgb}{0,0,0.8}
\definecolor{fhcolor}{rgb}{0,0.5,0}

\onlineid{0}

\vgtccategory{Research}

\vgtcinsertpkg

\title{Supporting Music Education through Visualizations of MIDI Recordings}

\author{Frank Heyen\thanks{e-mail: frank.heyen@visus.uni-stuttgart.de}\\ %
     \parbox{1.4in}{\scriptsize \centering VISUS,\\University of Stuttgart} %
\and Michael Sedlmair\thanks{e-mail: michael.sedlmair@visus.uni-stuttgart.de}\\ %
     \parbox{1.4in}{\scriptsize \centering VISUS,\\University of Stuttgart}}

\abstract{
Musicians mostly have to rely on their ears when they want to analyze what they play, for example to detect errors.
Since hearing is sequential, it is not possible to quickly grasp an overview over one or multiple recordings of a whole piece of music at once.
We therefore propose various visualizations that allow analyzing errors and stylistic variance.
Our current approach focuses on rhythm and uses MIDI data for simplicity.
}
\CCScatlist{
    \CCScatTwelve{Human-centered computing}{Information visu\-al\-iza\-tion}{}{}{}
    \CCScatTwelve{Applied computing}{Interactive learning environments}{}{}
}

\begin{document}

\firstsection{Introduction}

\maketitle

Our work focuses on investigating how visual data analysis can be leveraged in music education and training.
While several works exist on how to support static sheet music through visualization~\cite{miller2019augmenting, wattenberg2002arcdiagrams}, 
our focus is on visualizing actual recordings of notes that musicians play while learning their instrument. 
We specifically hypothesize that comparative visualization~\cite{gleicher2011visual} can be a viable approach for leveraging such data. 
In a traditional tutoring setup, for instance, we could record what both a student and a teacher are playing, and then use this data for comparative and easy to understand visualizations. For auto-didactic learning, we could compare actual played notes of a student to a ``ground truth'' piece of sheet music. 
Comparative visualization could then be used to help identify errors in playing and guide the process of future practice.

The data behind such an analysis can be represented in different formats, such as MusicXML~\cite{good2001musicxml} or MIDI (Musical Instrument Digital Interface) for abstract notes, or WAV (Waveform Audio File Format) for raw audio recordings.
To keep the concept simple, we decided to start with MIDI data instead of raw audio, since abstract notes are easier to interpret and visualize by a computer. 
While traditionally MIDI was closely connected to keyboards, many other instruments now natively support it, such as electronic drum kits, MIDI guitars, and even MIDI saxophones. 

In this work, we contribute initial design considerations and prototypical implementations for comparative MIDI visualizations.
The poster focuses on MIDI drum recordings and the task of analyzing rhythm in terms of consistency and errors.

Methodologically we are inspired by design study methodology~\cite{sedlmair2012design} and action research~\cite{hayes2011relationship}.
One of the authors taught guitar for seven years at a music school and is currently learning drums.

\section{Related Work}

Although there is diverse research regarding music visualization~\cite{khulusi2020survey} and visual musicology~\cite{miller2019framing}, there is not much usage of instrument-generated data for education yet.
Rocksmith\footnote{\url{https://rocksmith.ubisoft.com/}} and Synthesia\footnote{\url{https://synthesiagame.com/}} are examples for instrument education games.
Such games only give immediate or coarsely aggregated feedback and do not allow analyzing recordings on a per-note basis.
In terms of visual encoding, they are mostly limited to showing coarse scores or line charts. 
They do not show detailed feedback on a per-note basis, restricting the insight a user might gain.
There has been little work in actively leveraging visualization and visual analytics for this endeavor, which is the focus of our work.

Abstractly, our data can be viewed as a sequence of played notes. Analyzing temporal data is a core area in visualization research~\cite{aigner2011visualization},
and various approaches for visual sequence analysis  exist~\cite{luo2012EventRiver,sedlmair2011information,Wongsuphasawat2011lifeflow}. 
We aim at contributing the exploration of a new application area in this field.

\section{Design}

This sections provides details on the kinds of data we consider, as well as users and tasks we aim to support.

\paragraph{Data}

We focus on MIDI data, which represents music as abstract notes instead of audio.
This makes our design simpler and more reliable, but still works with a range of instruments.
We differentiate between two kinds of data: \emph{recordings} and \emph{ground truth}.
Recordings capture the MIDI output of an instrument, saving the played notes for later analysis, while the ground truth is the correct version of a piece of music that the users wants to learn.
MIDI files containing ground truth are abundant online and easy to create with existing tools.
% For example, a search for \emph{MIDI drum dataset} reveals a collection of 800,000 drum patterns\footnote{\url{https://www.reddit.com/r/WeAreTheMusicMakers/comments/3anwu8/the_drum_percussion_midi_archive_800k/}}.
Alternatively, a MIDI recording from a music teacher might be used.
Without access to a ground truth, for example when improvising or composing, some of our visualizations are still able to provide insight.

\paragraph{Users and Tasks}

Our target users are musicians who want to learn or improve their skills at playing an instrument.
We want to enable those users to see where they make errors such as hitting the wrong note or playing it slightly too early or too late.
To reduce effort and allow for temporal overviews, visualizations should allow analyzing multiple recordings at once, to better reveal consistent patterns.
For example, a drummer might find that he consistently lags behind on the bass drum.
This kind of consistent offset might even be on purpose, depending on playing style.

\paragraph{Visual Encodings}

\begin{figure}[htb]
 \centering
 \includegraphics[width=\columnwidth]{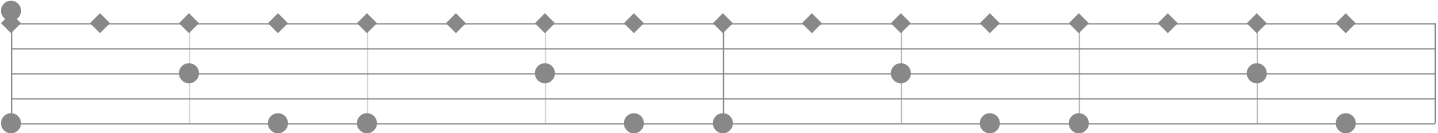}
 \caption{
 A simplified, notation-based representation of the ground truth drum pattern.
 This serves as basis for most of our visualizations.
 }
 \label{fig:ground_truth}
\end{figure}

In order to create intuitive visualizations that reduce the effort for learning and reading, we build upon standard music notation, such as scores and tablature.
We draw measures and beats with a size proportional to their duration in seconds to allow perceiving durations through pre-attentive processing (\autoref{fig:ground_truth}).

\begin{figure}[ht]
 \centering
 \includegraphics[width=\columnwidth]{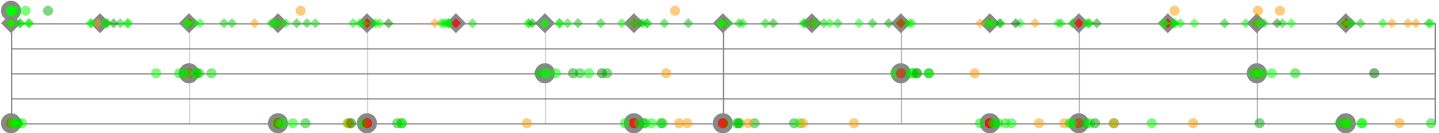}
 \caption{
 Recorded notes of multiple recordings (colored) drawn on top of the ground truth notes.
 Colors indicate \textcolor{red}{missing}, \textcolor{orange}{surplus}, and \textcolor{ForestGreen}{correct} notes.
 }
 \label{fig:error_colors}
\end{figure}

With a single recording, a naive approach of simply plotting all ground truth notes and recording notes works fine.
We use colors to encode the kind of error as follows: red for missing notes, yellow for surplus ones, and green for correct notes (\autoref{fig:error_colors}).
Notes are considered correct if they can be matched to a ground truth note, such that for each recording at most one recorded note is assigned to one ground truth note.
Correct notes are usually not perfectly timed; we use different shades of green to encode the timing error.
While this visualization has the advantage of showing error types, it does not scale to larger numbers of recordings, as drawing all notes leads to over-plotting.

\begin{figure}[ht]
 \centering
 \includegraphics[width=\columnwidth]{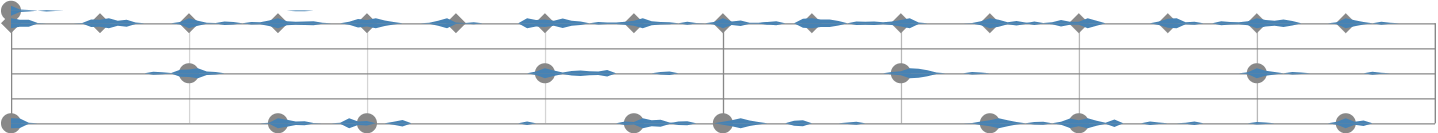}
 \caption{
 To reduce clutter, this visualization does not draw single notes but instead uses a density estimation area chart to show the temporal distribution of recorded notes.
 }
 \label{fig:density}
\end{figure}

A more scalable alternative is to draw one density estimation chart per pitch (\autoref{fig:density}).
In the case of drum patterns, this means one chart for each of its components such as the bass drum and cymbals.
For example, we take the start times of all hi-hat notes from all recordings and calculate the density estimation for them.
Through this aggregation, an arbitrary number of recordings can be shown at once to analyze general patterns such as syncopation.

\begin{figure}[ht]
 \centering
 \includegraphics[width=\columnwidth]{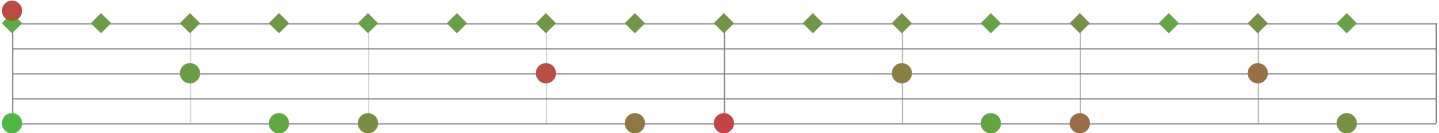}
 \caption{
 Drawing only the ground truth, colored by the average error over all recordings, creates a clear image that provides insight into where the problematic parts are.
 }
 \label{fig:gt_colored}
\end{figure}

The most concise encoding only draws the notes of the ground truth data (\autoref{fig:gt_colored}).
Each note gets colored by the average error of the recorded notes that are matched to it.
This allows users to quickly spot the most problematic parts of the piece and may serve as an overview to start a more thorough analysis from.

\begin{figure}[htp]
 \centering
 \includegraphics[width=\columnwidth]{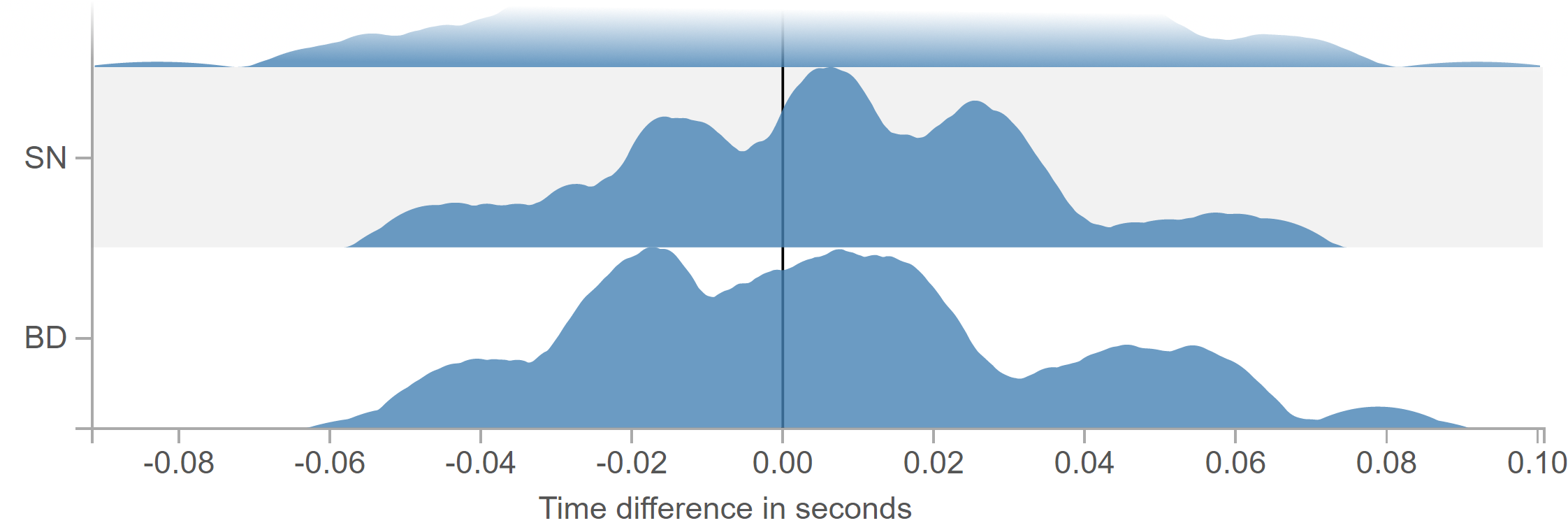}
 \caption{
 A time-aggregated view shows the distribution of time differences between played and matched correct notes for each drum kit component (here BD = bass drum, SN = snare).
 }
 \label{fig:perpitchsummary}
\end{figure}

Since different drum components are played differently, users might be interested in a summary for each component.
We implemented such a visualization that aggregates over the whole duration of a track.
With a density estimation area chart, it shows the distribution of the time differences between played and correct notes of all recordings in a compact fashion (\autoref{fig:perpitchsummary}).
This visualization could reveal that a user makes larger errors on the bass drum, which is played via foot, than components that are played with hands.

\section{Conclusion and Future Work}

We proposed five visual encodings for musical performance analysis using drum MIDI recordings.
Currently, we are in the process of integrating these visualizations into an interactive prototype that allows users to detect errors and analyze their style of playing.
In terms of interaction, we are implementing different mechanisms for selections to dynamically aggregate sub-sequences in the data.
Our current version already supports synthesized playback of both the ground truth and single recordings, other visual encodings, linking and brushing, and other features. 
At the heart, these are different versions of the above five representations though.
Qualitatively evaluating first versions of this prototype indicated that users might indeed profit from augmenting their auditory analysis with visual encodings. 
In the future, we also plan dedicated comparative studies to better understand the design space.

As we focus on MIDI data only, our approach is limited to instruments that are able to produce the required output.
We plan to extend our concept to different kinds of instruments that support MIDI, such as guitars with MIDI pickups, and will also look into incorporating instruments without native MIDI support.

\bibliographystyle{abbrv-doi-hyperref-narrow}

\bibliography{bib}
\end{document}